\documentclass{PoS} 
\usepackage{axodraw} 
\usepackage{pstricks,pst-plot,pst-node,
pst-coil}
\usepackage{graphicx}
\usepackage{amsmath}
\usepackage{amssymb}
\PoS{PoS(LAT2005)353} 

\title{Hadronic light-by-light scattering contribution 
to the muon $g-2$ from lattice QCD:\,Methodology} 

\ShortTitle{Hadronic light-by-light scattering contribution 
to the muon $g-2$ from lattice QCD} 
\author{\speaker{Masashi Hayakawa}\\ 
        Theoretical Physics Group, RIKEN, 
         Wako 2-1, Saitama 351-0198, Japan\\ 
        E-mail: \email{haya@riken.jp}} 

\author{Thomas Blum\\
        Physics Department, University of Connecticut, 
         Storrs, CT\ 06269-3046, USA\\ 
        RIKEN-BNL Research Center, Brookhaven National Laboratory, 
         Upton, New York 11973, USA\\ 
        E-mail: \email{tblum@phys.uconn.edu}} 

\author{Taku Izubuchi\\
        Kanazawa University, 
        Institute for Theoretical Physics, 
         Kanazawa 920-1192, Japan\\ 
        RIKEN-BNL Research Center, Brookhaven National Laboratory, 
         Upton, New York 11973, USA\\ 
        E-mail: \email{izubuchi@quark.phy.bnl.gov}} 

\author{Norikazu Yamada\\
        High Energy Accelerator Research Organization {\rm (}KEK{\rm )}, 
         Tsukuba, Ibaraki 305-0801, Japan\\ 
        The Graduate University for Advanced Studies, 
         Tsukuba, Ibaraki 305-0801, Japan\\ 
        E-mail: \email{norikazu.yamada@kek.jp}} 

\abstract{ 
 The hadronic light-by-light scattering contribution 
to the muon $g-2$ is the most troublesome component of 
its theoretical prediction; 
(1) it cannot be determined from the other measurable quantities, 
(2) the dimensional argument and the estimation based on hadronic models 
imply that the magnitude of this contribution may be comparable 
to the discrepancy between the standard model prediction 
and the experimental value. 
 The direct approach to evaluate 
the hadronic light-by-light scattering contribution 
requires the evaluation of the correlation function 
of {\it four} hadronic electromagnetic currents, 
and the summation of it
 over two independent four-momenta of off-shell photons, 
which is far from the reach of direct lattice simulation. 
 Here we propose an alternative method 
using combined (QCD + QED) lattice simulations 
to evaluate the hadronic light-by-light scattering contribution.  
}

\FullConference{XXIIIrd International Symposium on Lattice Field Theory\\
		 25-30 July 2005\\
		 Trinity College, Dublin, Ireland}

\begin{document}

%%%%%%%%%%%%%%%%%%%%%%%%%%%%%%%%%%%%%%%%%%%%%%%%%%%%%%%%%%%%%% 
%%\section{Introduction} 
%%\label{sec:introduction} 
%%%%%%%%%%%%%%%%%%%%%%%%%%%%%%%%%%%%%%%%%%%%%%%%%%%%%%%%%%%%%% 

 The muon anomalous magnetic moment ($g-2$) 
has been measured very accurately 
%with great accuracy 
%$\Delta a_\mu({\rm EXP})$ $= 6 \times 10^{-10}$ 
at BNL 
\cite{Bennett:2002jb} 
\begin{equation} 
 \Delta a_\mu({\rm EXP}) = 6 \times 10^{-10}. 
  \label{eq:BNL_accuracy}
\end{equation} 
 Theoretical interest in the muon $g-2$ stems from the fact 
that it can be used as a probe of unknown microscopic structures, 
such as supersymmetry with its breaking scale 
$m_S \sim {\cal O}(100)$ GeV and large $\tan \beta (\ge 10)$, 
and TeV scale gravity 
\cite{Arkani-Hamed:1998rs}, 
each of which can give the additional contribution  
$a_\mu({\rm new\ physics}) = {\cal O}(10 \sim 100) \times 10^{-10}$. 
 We recall that 
both of these two structures 
are the most promising possibilities 
to solve the hierarchy problem, 
i.e., the hierarchy between the electroweak scale 
and the GUT or Planck scale. 
 Thus, the currently available accuracy of $\Delta a_\mu({\rm EXP})$ 
can provide us with an opportunity to examine 
if the underlying structures ensuring such a hierarchy 
exist. 

 Table \ref{table:theory_exp} summarizes the current status 
of the muon $g-2$ \cite{Hertzog:2005mx}, 
% Table \ref{table:theory_exp} shows 
showing 
$2.7 \sigma$ deviation of $a_\mu({\rm EXP})$ 
from the standard model prediction $a_\mu({\rm SM})$ 
\cite{Bennett:2002jb,Hertzog:2005mx}. 
 However,  
the uncertainty in $a_\mu({\rm SM})$, 
which is attributed to the QCD contribution, 
must be reduced to elaborate such a discrepancy. 
% However, the theoretical uncertainty must be 
%reduced further to elaborate such a discrepancy. 
% From Table \ref{table:theory_exp}, 
%the uncertainty of $a_\mu({\rm SM})$ 
%is attributed to the QCD contribution. 
%% 
\begin{table}[hbt]  
 \begin{center} 
  \begin{tabular}{lrl} 
   \hline 
     & $a_\mu \times 10^{10}$ & $\Delta a_\mu \times 10^{10}$\\ 
   \hline 
   QED & $11\ 658\ 471.94$ & $0.14$ \\ 
%   \textcolor[named]{Red}{QCD} & \textcolor[named]{Red}{$695.4$} 
%    & \textcolor[named]{Red}{$7.3$} \\ 
   {\red QCD} & ${\red 695.4}$ & ${\red 7.3}$ \\ 
   Weak & $15.4$ & $0.22$ \\ 
   \hline 
%   Theory & $11\ 659\ 182.7$ & \textcolor[named]{Blue}{$7.4$} \\ 
   Theory & $11\ 659\ 182.7$ & ${\red 7.4}$ \\ 
   \hline 
   Experiment & $11\ 659\ 208\ \ $ & $6$ \\ 
   \hline 
   $a_\mu({\rm EXP}) - a_\mu({\rm SM})$ 
%    & \textcolor[named]{Red}{25.3} & 9.5 \\ 
    & {\red 25.3} & {\red 9.5} \\ 
   \hline  
  \end{tabular} 
  \caption{Comparison of $a_\mu({\rm SM})$ with $a_\mu({\rm EXP})$} 
  \label{table:theory_exp}
 \end{center} 
\end{table} 

 Table \ref{table:muong-2fromQCD} summarizes 
the QCD contribution to $a_\mu({\rm SM})$ 
as well as 
its various components 
which are relevant in considering $\Delta a_\mu({\rm EXP})$ 
in Eq.~(\ref{eq:BNL_accuracy}). 
 The QCD contribution starts at ${\cal O}(\alpha_{\rm em}^2)$ 
through the hadronic contribution 
to the vacuum polarization of the photon, 
which has been computed 
with lattice QCD 
\cite{Blum:2002ii}, 
and by using chiral perturbation theory and a vector meson model 
to fit the lattice data \cite{Aubin}. 
% Practically, 
%the ${\cal O}(\alpha_{\rm em}^2)$ hadronic vacuum polarization 
%contribution must be determined at the level of ${\cal O}(1)$ \%, 
%and is inferred by using the measurement of 
%$\sigma(e^+ e^- \rightarrow {\rm hadrons})(\sqrt{s})$ 
%for low $\sqrt{s}$. 
 The improvement of the accuracy of this contribution 
either relies on the further precise measurement 
of $\sigma(e^+ e^- \rightarrow {\rm hadrons})(\sqrt{s})$ 
%for low $\sqrt{s}$ 
\cite{Hagiwara:2003da}, 
%(under active debate \cite{Hagiwara:2003da}), 
or on the improvement of the lattice calculation 
with a statistical uncertainty less than 5 \% \cite{Aubin} 
(but with less well-known systematic uncertainty).  
\begin{table}[hbt] 
%  \caption{QCD contribution to the muon $g-2$}
 \begin{center} 
  \begin{tabular}{lrl} 
   \hline 
     & $a_\mu \times 10^{10}$ & $\Delta a_\mu \times 10^{10}$\\ 
   \hline 
   hadronic vacuum polarization 
     (${\cal O}(\alpha_{\rm em}^{{\blue 2}})$) 
    & $693.4$ & $6.4$ \\ 
   hadronic vacuum polarization (${\cal O}(\alpha_{\rm em}^{{\blue 3}})$) 
    & $-10.0$ & $0.6$ \\ 
   {\red hadronic light-by-light} 
    (${\cal O}(\alpha_{\rm em}^{{\blue 3}})$) 
    & ${\red 12.0}$ & ${\red 3.5}$ \\ 
   \hline 
   Total QCD & $695.4$ & $7.3$ \\ 
   \hline  
  \end{tabular} 
  \caption{QCD contribution to the muon $g-2$  \cite{Hertzog:2005mx}} 
  \label{table:muong-2fromQCD} 
 \end{center} 
\end{table} 

\begin{figure}[htb] 
 \begin{picture}(200,160)(-120,0) 
%% muon line 
  \SetWidth{1.2} 
  \SetColor{Blue} 
  \ArrowLine(-50,0)(-100,0) 
  \ArrowLine(0,0)(-50,0) 
  \ArrowLine(50,0)(0,0) 
  \ArrowLine(100,0)(50,0) 
  \SetColor{Black} 
  \Text(80,-5)[t]{$\mu$} 
%% Hadronic blob 
  \COval(0,80)(50,50)(45){Gray}{Gray} 
%% Photon lines 
  \SetColor{Red} 
  \Photon(-50,0)(-40,50){3}{4}  
  \Photon(0,0)(0,30){3}{3}  
  \Photon(50,0)(40,50){3}{4} 
  \Photon(0,130)(0,160){3}{2.5} 
  \SetColor{Black} 
%% vertices on blob 
  \SetColor{Green}  
  \Vertex(-50,0){3} 
  \Vertex(-0,0){3} 
  \Vertex(50,0){3} 
  \Vertex(0,130){3} 
%% vertices on muon line 
  \SetColor{Green}  
  \Vertex(-40,50){3} 
  \Vertex(0,30){3} 
  \Vertex(40,50){3} 
%% explanation on blob part 
  \Text(150,100)[t]{elastic scattering amplitude} 
  \Text(150,85)[t]{of two photons by QCD} 
%% flow of loop momenta 
  \SetColor{Magenta} 
  \ArrowArc(22,15)(13,-40,220)  
  \ArrowArc(-25,15)(13,-40,220) 
%% explanation on loop momenta 
  \Text(22,20)[t]{$l_1$} 
  \Text(-25,20)[t]{$l_2$} 
 \end{picture} 
 \caption{hadronic light-by-light scattering contribution 
          to the muon $g-2$} 
  \label{fig:hadronic_light-by-light}
\end{figure}
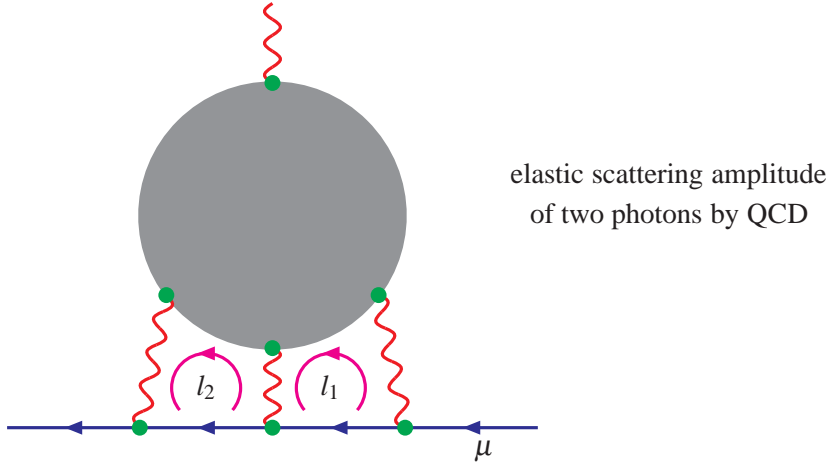

 Our target here is another type of 
%${\cal O}(\alpha_{\rm em}^3)$ 
QCD contribution, 
namely 
{\it the hadronic light-by-light scattering {\rm (}h-lbl{\rm )} contribution} 
to the muon $g-2$ shown in Fig.~\ref{fig:hadronic_light-by-light}, 
which also affects the physical interpretation of the muon $g-2$ 
in viewing Table \ref{table:muong-2fromQCD}. 
 The diagram in Fig.~\ref{fig:hadronic_light-by-light} 
arises through the elastic scattering amplitude 
of two (off-shell) photons {\it by QCD} (the blob). 
 To date, this contribution could be estimated 
by purely theoretical calculation. 
 So far, it has been calculated only based 
on the hadronic picture 
\cite{hadronic_lbl,pseudoscalar}. 
 Thus the first principle calculation based on lattice QCD is 
particularly desirable. 

 The diagram in Fig.~\ref{fig:hadronic_light-by-light} 
evokes the following naive approach; 
we calculate {\it repeatedly} 
the correlation function of 
{\it four} hadronic electromagnetic currents 
%$j_\mu(x)$
by lattice QCD 
{\it with respect to two independent four-momenta} 
$l_1,\,l_2$ of off-shell photons, 
and integrate it over $l_1,\,l_2$. 
 Such a task is too difficult to accomplish 
with use of supercomputers available in the foreseeable future. 
% Such a task is too hard 
%to be accomplished even 
%with use of the future supercomputers 
%available within the next few decades. 
% The calculation requires innovation so that it can be carried out 
%within a practical amount of computation. 

%%%%%%%%%%%%%%%%%%%%%%%%%%%%%%%%%%%%%%%%%%%%%%%%%%%%%%%%%%% 
%%\section{Methodology} 
%%\label{sec:methodlogy} 
%%%%%%%%%%%%%%%%%%%%%%%%%%%%%%%%%%%%%%%%%%%%%%%%%%%%%%%%%%% 
\vspace{0.3cm} 
 Here we propose a practical method to calculate 
the h-lbl contribution 
by using the lattice (QCD + {\it QED}) simulation; 
%aiming to determine its sign as well as its order of magnitude; 
we compute 
\begin{eqnarray} 
 && 
 \left< 
  \begin{picture}(70,40)(-5,20) 
   %% quark loop 
    \SetWidth{1.2} 
    \ArrowArc(30,40)(15,90,180) 
    \ArrowArc(30,40)(15,180,270) 
    \ArrowArc(30,40)(15,270,360) 
    \ArrowArc(30,40)(15,0,90) 
    \Text(30,43)[t]{\scriptsize quark} 
   %% photon lines 
    \SetWidth{1.2} 
    \SetColor{Red} 
    \Photon(30,55)(30,65){1.5}{2.5} 
    \Photon(30,25)(30,5){1.5}{4} 
   %% vertices 
    \SetColor{Green} 
    \Vertex(30,55){2} 
    \Vertex(30,25){2} 
   %% muon lines 
    \SetWidth{1.2} 
    \SetColor{Blue} 
    \ArrowLine(60,5)(30,5) 
    \ArrowLine(30,5)(0,5) 
   %% vertex 
    \SetColor{Green} 
    \Vertex(30,5){2}
  \end{picture} 
 \right>_{{\rm QCD} + {\rm quenched\ QED}_A} 
%  \nonumber \\ 
%  \nonumber \\ 
%  \nonumber \\ 
% && 
 %% subtraction term 
% - \quad 
 - 
 \left< 
  \begin{picture}(70,20)(20,0) 
   %% quark loop 
   \SetWidth{1.2} 
   \ArrowArc(55,0)(15,90,180) 
   \ArrowArc(55,0)(15,180,270) 
   \ArrowArc(55,0)(15,270,360) 
   \ArrowArc(55,0)(15,0,90) 
   \Text(55,3)[t]{\scriptsize quark} 
   %% photon lines 
   \SetWidth{1.2} 
   \SetColor{Red} 
   \Photon(55,15)(55,25){1.5}{2.5} 
   \Photon(55,-15)(55,-35){1.5}{4} 
   %% vertices 
   \SetColor{Green} 
   \Vertex(55,15){2} 
   \Vertex(55,-15){2} 
  \end{picture} 
 \right>_{{\rm QCD} + {\rm quenched\ QED}_B} 
  \nonumber \\ 
 && 
 \qquad 
 \qquad \qquad \qquad \qquad \qquad \qquad 
 \qquad \,  
  \left< 
   \begin{picture}(70,20)(-5,0) 
    %% muon lines 
    \SetWidth{1.2} 
    \SetColor{Blue} 
    \ArrowLine(60,5)(30,5) 
    \ArrowLine(30,5)(0,5) 
    %% vertex 
    \SetColor{Green} 
    \Vertex(30,5){2} 
   \end{picture} 
  \right>_{{\rm quenched\ QED}_A} \, , 
   \label{eq:method_figure} 
\end{eqnarray} 
amputate the external muon lines, 
and project the magnetic form factor, 
and divide by the factor $3$. 
 In Eq.~(\ref{eq:method_figure}) 
the red line denotes the free photon propagator 
$D_{\mu\nu}(x,\,y)$ 
in the non-compact lattice QED solved in an appropriate gauge 
fixing condition. 
 The black line denotes the full quark propagator 
$S_f(x,\,y;U,\,u)$ 
for a given set of $SU(3)_C$ gauge configuration 
$\left\{U_{x,\,\mu}\right\}$ 
and $U(1)_{\rm em}$ gauge configuration $\left\{u_{x,\,\mu}\right\}$, 
where the sum over relevant flavors $f$ is implicitly assumed. 
 The blue line represents the full muon propagator 
$s(x,\,y;\,u)$. 
 The average $\left<,\,\right>$ above means 
the one over the unquenched $SU(3)_C$ gauge configurations 
and/or the quenched $U(1)_{\rm em}$ gauge configurations 
\footnote{ 
 For the unquenched QCD plus quenched QED 
to respect the gauge invariance of QED, 
the electromagnetic charges of sea quarks are assumed to be zero.  
} as specified by the subscript attached to it. 
 Since two statistically independent averages 
over $U(1)_{\rm em}$ gauge configurations appear 
in the second term, they are distinguished by the labels $A$, $B$. 

% To show that the magnetic component of Eq.~(\ref{eq:method_figure}) 
%is related to our target, 
 To explain the mechanism underlying our method, 
let us look at the first term of Eq.~(\ref{eq:method_figure}) 
{\it perturbatively with respect to QED}. 
 Its magnetic components up to ${\cal O}(\alpha_{\rm em}^3)$ 
consist of 
%\footnote{ 
% Consideration here also persists 
%even in the presence of the diagrams 
%arising from the lattice artifact interactions, 
%which are not shown in this paper. 
%}
%% 
\begin{eqnarray} 
 %% O(alpha^2) correction 
 && {\cal O}(\alpha_{\rm em}^{2}) 
    + {\cal O}(\alpha_{\rm em}^{3}) \qquad \qquad 
 \left< 
  \begin{picture}(80,20)(20,0) 
   %% quark loop 
   \SetWidth{1.2} 
   \ArrowArc(60,0)(15,90,180) 
   \ArrowArc(60,0)(15,180,270) 
   \ArrowArc(60,0)(15,270,360) 
   \ArrowArc(60,0)(15,0,90) 
   \Text(60,3)[t]{\scriptsize quark}
   %% photon lines 
   \SetColor{Red} 
   \SetWidth{1.2} 
   \Photon(60,15)(60,25){1.5}{2.5} 
   \Photon(60,-15)(60,-35){1.5}{4} 
   %% vertices 
   \SetColor{Green} 
   \Vertex(60,15){2} 
   \Vertex(60,-15){2} 
  \end{picture} 
 \right>_{\rm QCD} 
  \nonumber \\ 
 && \qquad \qquad \qquad \qquad \qquad \quad \ \  
  \left< 
  \begin{picture}(80,20)(0,0) 
   %% muon lines 
   \SetWidth{1.2} 
   \SetColor{Blue} 
   \ArrowLine(70,3)(50,3) 
   \Line(50,3)(40,3) 
   \Line(40,3)(30,3) 
   \ArrowLine(30,3)(10,3) 
   %% photon lines 
   \SetWidth{1.2} 
   \SetColor{Red} 
   \PhotonArc(40,3)(10,180,360){1.5}{6} 
   %% vertices 
   \SetColor{Green} 
   \Vertex(50,3){2} 
   \Vertex(40,3){2} 
   \Vertex(30,3){2} 
  \end{picture} 
 \right>_{\rm 1 + {\cal O}(\alpha_{\rm em})}
 \nonumber \\ 
 \nonumber \\ 
 \nonumber \\ 
 && {\cal O}(\alpha_{\rm em}^{3}) \quad 
 %% light-by-light 
  \left< 
   \begin{picture}(80,20)(20,0) 
    %% quark loop 
    \SetWidth{1.2} 
    \ArrowArc(60,0)(15,90,270) 
    \ArrowArc(60,0)(15,270,450) 
    \Text(60,3)[t]{\scriptsize quark} 
    %% photon lines 
    \SetColor{Red} 
    \SetWidth{1.2} 
    \Photon(60,15)(60,25){1.5}{2.5} 
    \Photon(50,-10)(43,-35){1.5}{5} 
    \Photon(60,-15)(60,-35){1.5}{4} 
    \Photon(70,-10)(77,-35){1.5}{5} 
    %% vertices 
    \SetColor{Green} 
    \Vertex(60,15){2} 
    \Vertex(50,-10){2} 
    \Vertex(60,-15){2} 
    \Vertex(70,-10){2}
   \end{picture} 
  \right>_{\rm QCD} 
  \quad  
  %% O(alpha) correction to quark loop 
  \left< 
   \begin{picture}(80,20)(20,0) 
   %% quark loop 
    \SetWidth{1.2} 
    \ArrowArc(60,0)(15,90,180) 
    \ArrowArc(60,0)(15,180,270) 
    \ArrowArc(60,0)(15,270,360) 
    \ArrowArc(60,0)(15,0,90) 
    \Text(60,3)[t]{\scriptsize quark}
   %% photon lines 
    \SetColor{Red} 
    \SetWidth{1.2} 
    \Photon(60,15)(60,25){1.5}{2.5} 
    \Photon(60,-15)(60,-35){1.5}{4} 
   %% vertices 
    \SetColor{Green} 
    \Vertex(60,15){2} 
    \Vertex(60,-15){2} 
   \end{picture} 
  \right>_{{\rm QCD} \times {\cal O}(\alpha_{\rm em})}
 \nonumber \\ 
 && \qquad \qquad  
 %% light-by-light 
  \begin{picture}(80,20)(20,0) 
   %% muon line  
   \SetWidth{1.2} 
   \SetColor{Blue} 
   \ArrowLine(107,3)(89,3) 
   \Line(89,3)(72,3) 
   \Line(72,3)(55,3) 
   \ArrowLine(55,3)(37,3) 
   %% vertices 
   \SetColor{Green} 
   \Vertex(89,3){2} 
   \Vertex(72,3){2} 
   \Vertex(55,3){2} 
  \end{picture} 
  \qquad \qquad \ \   
  %% O(alpha) correction to quark loop 
  \begin{picture}(80,20)(20,0) 
  %% muon lines  
  \SetWidth{1.2} 
  \SetColor{Blue} 
  \ArrowLine(100,3)(80,3) 
  \Line(80,3)(70,3) 
  \Line(70,3)(60,3) 
  \ArrowLine(60,3)(40,3) 
  %% photon lines 
  \SetWidth{1.2} 
  \SetColor{Red} 
  \PhotonArc(70,3)(10,180,360){1.5}{6} 
  %% vertices 
  \SetColor{Green} 
  \Vertex(80,3){2} 
  \Vertex(70,3){2} 
  \Vertex(60,3){2} 
  \end{picture} 
   \, . 
   \label{eq:O(alpha)-expansion_first-term} 
%  \nonumber \\ 
%  \nonumber \\ 
% 
% && 
% \qquad \qquad 
% %% correction to the muon line 
% \left< 
%  \begin{picture}(80,20)(20,0) 
%   %% quark loop 
%   \SetWidth{1.2} 
%   \ArrowArc(60,0)(15,90,180) 
%   \ArrowArc(60,0)(15,180,270) 
%   \ArrowArc(60,0)(15,270,360) 
%   \ArrowArc(60,0)(15,0,90) 
%   %% photon lines 
%   \SetColor{Red} 
%   \SetWidth{1.2} 
%   \Photon(60,15)(60,25){1.5}{2.5} 
%   \Photon(60,-15)(60,-35){1.5}{4} 
%   \Text(60,3)[t]{\tiny quark}
%   %% vertices 
%   \SetColor{Green} 
%   \Vertex(60,15){2} 
%   \Vertex(60,-15){2} 
%  \end{picture} 
% \right>_{\rm QCD} 
\end{eqnarray} 
 The diagram in the first line 
gives the ${\cal O}(\alpha_{\rm em}^2)$-contribution. 
 The ${\cal O}(\alpha_{\rm em})$-corrections 
to its muon part 
and to its quark part  
induce the ${\cal O}(\alpha_{\rm em}^3)$-contributions 
shown in the first line 
and in the right diagram on the second line 
respectively. 
 We recall that the QED gauge configurations 
in the first term of Eq.~(\ref{eq:method_figure}) 
are commonly shared by the quark part and the muon part. 
 Hence, the photons can be exchanged between the two parts. 
 As a consequence, the left diagram in the second line 
of Eq.~(\ref{eq:O(alpha)-expansion_first-term}) is induced 
at ${\cal O}(\alpha_{\rm em}^3)$, 
which takes the form of our target, 
Fig.~\ref{fig:hadronic_light-by-light}. 
 Alternatively, 
the quark and muon parts 
in the first and third diagrams  
in Eq.~(\ref{eq:O(alpha)-expansion_first-term}) 
are connected only by a single photon attached a priori. 
 The second term in Eq.~(\ref{eq:method_figure}) 
also contains those extra diagrams. 
 Thus, by subtracting the second term from the first term, 
we may extract the h-lbl contribution. 
%\footnote{ 
% The subtraction of the second term from the first term 
%in Eq.~(\ref{eq:method_figure}) 
%is also motivated to 
%remove the overall ultraviolet singularity 
%in the hadronic correction to the vacuum polarization 
%which arises when the quark loop shrinks to a point. 
%}. 

 The quantities evaluated in our method (\ref{eq:method_figure}) 
are constructed from {\it two} currents for both terms, 
which are surely less noisy than 
the case of {\it four} currents encountered in the naive approach. 
 Amazingly, 
the only difference between the first and second terms  
of Eq.~(\ref{eq:method_figure}) 
is the ways averaging over the $U(1)_{\rm em}$ gauge configurations. 
 The h-lbl contribution 
should thus emerge as such a subtle difference in averaging procedure. 

 For our method to work efficiently, it is important to implement 
the cancellation of ${\cal O}(\alpha_{\rm em}^2)$-term 
in Eq.~(\ref{eq:O(alpha)-expansion_first-term}) 
without too much statistics.
 Sharing the same ensemble of 
$U(1)_{\rm em}$ gauge configurations 
between the first term and the ${\rm QED}_A$ 
part of the second term is one such method 
since the ${\cal O}(\alpha_{\rm em}^2)$-correction 
to the muon vertex in both terms 
will be highly correlated. 
 Furthermore, we organize Eq.~(\ref{eq:method_figure}) in the manner; 
%\footnote{ 
% To remove the vertex corrections on the muon line, 
%the wave function renormalization constant $Z_\mu$ of the muon 
%should be multiplied as an overall factor. 
%}; 
%% 
\begin{eqnarray} 
 && 
 H_\mu(t_F,\,x_c,\,t_I) 
  \nonumber \\ 
 && \  
 \equiv 
 \frac{Z_\mu}{3} 
 \left< 
  \left< 
  \sum_{x,\,y} 
   \sum_{\vec{x}_F} 
    e^{-i \vec{p}_F \cdot \vec{x}_F}\, 
    s(x_F,\,x;\,u_A)\, 
   %\times 
   e \gamma_\rho %j_\rho(x,\,y^\prime;\,u_A) 
%  \right. 
% \right. 
%  \nonumber \\ 
% && \quad 
% \qquad \qquad 
%  \left. 
%   \times 
   \sum_{\vec{x}_I} 
    e^{i \vec{p}_I \cdot \vec{x}_I}\, 
    s(x,\,x_I;\,u_A) 
  \right. 
 \right. 
  \nonumber \\ 
 && \qquad \ 
 \times 
 D_{\rho \lambda}(x,\,y) 
  \nonumber \\ 
 && \qquad  
 \times 
% \left< 
  \sum_f Q_f^2 
  \left( 
   {\rm tr} 
   \left( 
    \gamma_\mu %\j_\mu(x_c,\,x;\,U,\,u_A)\, 
    S_f(x_c,\,y;\,U,\,u_A)\, 
%   \right. 
%  \right. 
%% \right. 
%  \nonumber \\ 
% && \quad \qquad \qquad 
%  \left. 
%   \left. 
%    \times 
    e \gamma_\lambda %j_\lambda(x^\prime,\,x^{\prime\prime};\,U,\,u_A)\,  
    S_f(y,\,x_c;\,U,\,u_A) 
   \right) 
  \right. 
  \nonumber \\ 
 && \quad \qquad \qquad \ \ 
 \left. 
 \left. 
  \left. 
   - 
   \left< 
    {\rm tr} 
    \left( 
     \gamma_\mu %j_\mu(x_c,\,x;\,U,\,u_B)\, 
     S_f(x_c,\,y;\,U,\,u_B) 
%    \right. \, 
%   \right. 
%  \right. 
% \right. 
%  \nonumber \\ 
% && \quad \qquad \qquad \quad 
% \left. 
% \left. 
%  \left. 
%   \left. 
%    \left. 
%     \times 
     e \gamma_\mu %j_\lambda(x^\prime,\,x^{\prime\prime};\,U,\,u_B)\,  
     S_f(y,\,x_c;\,U,\,u_B) 
    \right) 
   \right>_{u_B} 
  \right) 
% \right>_U 
 \right>_{u_A} 
\right>_U 
  \, , 
  \label{eq:method_equation}
\end{eqnarray} 
where 
%``{\rm tr}'' denotes the sum over color and spinor indices, and 
the local form for the electromagnetic currents 
is assumed to avoid the complicate expression, 
and the wave function renormalization constant $Z_\mu$ of the muon 
is multiplied to remove the UV singularities in the vertex function 
of the muon. 
%$j_\mu(x,\,y;\,U,\,u)$ 
%($j_\mu(x,\,y;\,u)$) $ \simeq \gamma_\mu\,\delta_{x,\,y}$
%is obtained from the conserved electromagnetic current 
%by splitting off the quark (muon) fields, 
 The ordering of the averages 
in Eq.~(\ref{eq:method_equation}) is organized in such a way 
that the cancellation of ${\cal O}(\alpha_{\rm em}^2)$-''components'' 
can be realized as accurately as possible 
at the level of each configuration $(U,\,u_A)$. 
 The nested sum for $x,\,y$ could be efficiently carried out 
%using fast Fourier transformation. 
by working in the momentum space, 
where the Fourier transform of $D(x,\,y)$ is known exactly. 
  
%%%%%%%%%%%%%%%%%%%%%%%%%%%%%%%%%%%%%%%%%%%%%%%%%%%%%%%%%%%%%%%%%% 
%%\section{Discussion} 
%%\label{sec:discussion} 
%%%%%%%%%%%%%%%%%%%%%%%%%%%%%%%%%%%%%%%%%%%%%%%%%%%%%%%%%%%%%%%%%% 
\vspace{0.3cm} 
% Since our method treats 
%QED nonperturbatively, 
%Eq.~(\ref{eq:method_equation}) 
%contains ${\cal O}(\alpha_{\rm em}^4)$-terms in general.  
 The observation done around
 Eq.~(\ref{eq:O(alpha)-expansion_first-term}) 
relies on the QED perturbation, 
while the simulation treats QED nonperturbatively. 
 Thus, for our method to work, it is necessary that 
(a) the simulation can be done at the small coupling constant 
$\alpha_{\rm em}$, possibly at its physical value 
$\alpha_{\rm em} = \frac{1}{137}$, 
and that 
(b) the results for the weak coupling allow 
perturbative interpretation. 
 As for (a), 
the {\it quenched non-compact} QED allows us to 
get the uncorrelated $U(1)_{\rm em}$ gauge configurations 
efficiently \cite{Duncan:1996xy}. 
 As for (b), 
the lattice (QCD + QED) study of 
the electromagnetic splitting of the pion masses 
demonstrates that the splitting is perturbative 
%($\propto \alpha_{\rm em}$) 
\cite{Yamada}, 
confidence that our method for the h-lbl contribution 
will work. 

 So far we have focused on the diagram 
consisting of a single quark loop 
with electromagnetic currents inserted. 
 In general, Fig.~\ref{fig:hadronic_light-by-light} 
contains two, three or four quark loops, 
where each quark loop contains at least one current. 
 In the quenched approximation to QED, 
they must be computed separately. 
%\footnote{  
 To illustrate implication of multi-quark loop diagrams, 
let us consider the charged pseudoscalar meson contribution 
\cite{hadronic_lbl} 
but with all electromagnetic currents put on its valence quarks, 
for simplicity. 
 Three types of two quark loop diagrams provide 
the meson diagrams where the currents lie on 
both of two valence quarks, 
while the single quark loop diagram can provide 
the diagram where all the currents lie on 
one of two valence quarks. 
%}. 
% Although we also have developed the optimized ways 
%to compute those multi-quark loop contributions, 
%the calculation will be surely harder 
%than the one for the single quark loop diagram. 
 As is well-known, 
the calculation of multi-quark loop contributions 
(disconnected diagrams) is difficult in lattice calculations, 
and will be surely harder than the single quark loop diagram. 
 Actually, the multi-quark loop diagrams 
vanish in the exact flavor $SU(3)$ limit, 
except for a two quark loop diagram 
where four currents distribute over two quark loops evenly. 
 It is an important issue 
to examine if they are subdominant and safely neglected. 

\vspace{0.3cm} 
%\acknowledgments{
 We thank RIKEN, Brookhaven National Laboratory
and the U.~S.~Department of Energy
for providing the facilities and hospitality
where this work was done. 
 M.~H.~'s work is supported in part by Grant-in-Aid 
for Scientific Research (15740173) in Japan. 
%}

%%%%%%%%%%%%%%%%%%%%%%%%%%%%%%%%%%%%%%%%%%%%%%%%%%%%%%%%%%%%%%%%%% 
%%%%%%%                  REFERENCES                       %%%%%%%% 
%%%%%%%%%%%%%%%%%%%%%%%%%%%%%%%%%%%%%%%%%%%%%%%%%%%%%%%%%%%%%%%%%% 
%% 
 
%% 

%%%%%%% 
%%%%%%% 
%%%%%%% 
%%%%%%% 
%%%%%%% 
%%%%%%% 
%%%%%%% 
%%%%%%% 

\begin{thebibliography}{99} 
% 
\bibitem{Bennett:2002jb}
  G.~W.~Bennett {\it et al.}  [Muon $g-2$ Collaboration], 
  \emph{Measurement of the positive muon anomalous magnetic moment 
        to 0.7-ppm},
  \emph{Phys.\ Rev.\ Lett.}\ {\bf 89} (2002) 101804, 
  \emph{Erratum-ibid.}\ {\bf 89} (2002) 129903 
  [{\tt hep-ex/0208001}]; %\\ 
  %%CITATION = HEP-EX 0208001;%% 
%% 
%\bibitem{Bennett:2004pv}
% G.~W.~Bennett {\it et al.}  [Muon g-2 Collaboration],
  \emph{Measurement of the negative muon anomalous magnetic moment 
        to 0.7-ppm},
  \emph{Phys.\ Rev.\ Lett.}\ {\bf 92} (2004) 161802  
  [{\tt hep-ex/0401008}]. 
  %%CITATION = HEP-EX 0401008;%% 
% 
 \bibitem{Arkani-Hamed:1998rs} 
  N.~Arkani-Hamed, S.~Dimopoulos and G.~R.~Dvali, 
   \emph{The hierarchy problem and new dimensions at a millimeter}, 
   \emph{Phys.\ Lett.} {\bf B\ 429} (1998) 263 
   [{\tt hep-ph/9803315}]. 
 %%CITATION = HEP-PH 9803315;%%
% 
 \bibitem{Hertzog:2005mx}
  D.~W.~Hertzog [E821 Collaboration],
   \emph{Measurement of the muon anomaly 
         to high and even higher precision}, 
   \emph{Nucl.\ Phys.\ Proc.\ Suppl.}\ {\bf 144} (2005) 191 
  [{\tt hep-ex/0501053}]. 
 %%CITATION = HEP-EX 0501053;%%
% 
 \bibitem{Blum:2002ii}
 T.~Blum,
   \emph{Lattice calculation of the lowest order 
         hadronic contribution 
         to the muon anomalous magnetic moment}, 
   \emph{Phys.\ Rev.\ Lett.}\ {\bf 91} (2003) 052001 
   [{\tt hep-lat/0212018}]; \\ 
  %%CITATION = HEP-LAT 0212018;%% 
 % 
 M.~Gockeler, R.~Horsley, W.~Kurzinger, D.~Pleiter, P.~E.~L.~Rakow 
 and G.~Schierholz [QCDSF Collaboration],
  \emph{Vacuum polarisation and hadronic contribution 
        to muon $g-2$ from lattice QCD}, 
  \emph{Nucl.\ Phys.} {\bf B\ 688} (2004) 135
  [{\tt hep-lat/0312032}]. 
  %%CITATION = HEP-LAT 0312032;%%
% 
 \bibitem{Aubin} 
  C.~Aubin and T.~Blum, 
   \emph{Lowest order hadronic contribution to the muon $g-2$}, 
   in proceedings of 
   \emph{XXIIIrd International Symposium on Lattice Field Theory}. 
% 
 \bibitem{Hagiwara:2003da}
  K.~Hagiwara, A.~D.~Martin, D.~Nomura and T.~Teubner,
   \emph{Predictions for $g-2$ of the muon and 
         $\alpha({\rm QED})(M_Z^2)$}, 
  \emph{Phys.\ Rev.} {\bf D\ 69} (2004) 093003 
  [{\tt hep-ph/0312250}]; \\ 
  %%CITATION = HEP-PH 0312250;%% 
% 
  M.~Davier, S.~Eidelman, A.~Hocker and Z.~Zhang,
   \emph{Updated estimate of the muon magnetic moment 
         using revised results from $e^+ e^-$ annihilation}, 
   \emph{Eur.\ Phys.\ J.} {\bf C 31} (2003) 503 
  [{\tt hep-ph/0308213}]. 
  %%CITATION = HEP-PH 0308213;%% 
% 
 \bibitem{hadronic_lbl} 
  M.~Hayakawa, T.~Kinoshita and A.~I.~Sanda,
   \emph{Hadronic light by light scattering effect 
         on muon $g-2$}, 
   \emph{Phys.\ Rev.\ Lett.}\ {\bf 75} (1995) 790
   [{\tt hep-ph/9503463}]; %\\ 
  %%CITATION = HEP-PH 9503463;%% 
%
   \emph{Hadronic Light-by-light Scattering Contribution 
         to Muon $g - 2$}, 
   \emph{Phys.\ Rev.} {\bf D\ 54} (1996) 3137
   [{\tt hep-ph/9601310}]; \\ 
  %%CITATION = HEP-PH 9601310;%% 
% 
  J.~Bijnens, E.~Pallante and J.~Prades, 
   \emph{Hadronic light by light contributions 
         to the muon $g-2$ in the large $N_C$ limit}, 
   \emph{Phys.\ Rev.\ Lett.}\ {\bf 75} (1995) 1447, 
   \emph{Erratum-ibid.}\ {\bf 75} (1995) 3781 
   [{\tt hep-ph/9505251}]; %\\ 
  %%CITATION = HEP-PH 9505251;%% 
% 
   \emph{Analysis of the Hadronic Light-by-Light Contributions 
        to the Muon $g-2$}, 
   \emph{Nucl.\ Phys.} {\bf B\ 474} (1996) 379
   [{\tt hep-ph/9511388}]. \\ 
  %%CITATION = HEP-PH 9511388;%%
% 
 \bibitem{pseudoscalar} 
 M.~Hayakawa and T.~Kinoshita,
  \emph{Pseudoscalar pole terms 
        in the hadronic light-by-light scattering contribution 
        to muon $g-2$}, 
  \emph{Phys.\ Rev.} {\bf D\ 57} (1998) 465, 
  \emph{Erratum-ibid.} {\bf D\ 66} (2002) 019902 
  [{\tt hep-ph/9708227}]; \\ 
  %%CITATION = HEP-PH 9708227;%%
% 
  M.~Knecht and A.~Nyffeler,
   \emph{Hadronic light-by-light corrections to the muon $g-2$: 
         The pion-pole contribution}, 
  \emph{Phys.\ Rev.} {\bf D\ 65} (2002) 073034 
  [{\tt hep-ph/0111058}]; \\ 
  %%CITATION = HEP-PH 0111058;%% 
% 
 M.~Knecht, A.~Nyffeler, M.~Perrottet and E.~De Rafael,
  \emph{Hadronic light-by-light scattering contribution 
        to the muon $g-2$: An effective field theory approach}, 
  \emph{Phys.\ Rev.\ Lett.}\ {\bf 88} (2002) 071802
  [{\tt hep-ph/0111059}]; \\ 
  %%CITATION = HEP-PH 0111059;%% 
% 
 M.~Hayakawa and T.~Kinoshita,
  \emph{Comment on the sign 
        of the pseudoscalar pole contribution to the muon $g-2$}, 
  hep-ph/0112102. \\ 
  %%CITATION = HEP-PH 0112102;%%
%  
 I.~Blokland, A.~Czarnecki and K.~Melnikov,
  \emph{Pion pole contribution 
        to hadronic light-by-light scattering 
        and muon anomalous magnetic moment}, 
  \emph{Phys.\ Rev.\ Lett.}\ {\bf 88} (2002) 071803
  [{\tt hep-ph/0112117}]; \\ 
  %%CITATION = HEP-PH 0112117;%% 
% 
 J.~Bijnens, E.~Pallante and J.~Prades,
  \emph{Comment on the pion pole part 
        of the light-by-light contribution to the muon $g-2$}, 
  \emph{Nucl.\ Phys.} {\bf B\ 626} (2002) 410
  [{\tt hep-ph/0112255}]; \\ 
  %%CITATION = HEP-PH 0112255;%% 
% 
 M.~Ramsey-Musolf and M.~B.~Wise,
  \emph{Hadronic light-by-light contribution 
        to muon $g-2$ in chiral perturbation theory}, 
  \emph{Phys.\ Rev.\ Lett.}\  {\bf 89} (2002) 041601 
  [{\tt hep-ph/0201297}]; \\ 
  %%CITATION = HEP-PH 0201297;%%
% 
 K.~Melnikov and A.~Vainshtein,
  \emph{Hadronic light-by-light scattering contribution 
        to the muon anomalous magnetic moment revisited}, 
  \emph{Phys.\ Rev.} {\bf D\ 70} (2004) 113006
  [{\tt hep-ph/0312226}]. 
  %%CITATION = HEP-PH 0312226;%% 
% 
 \bibitem{Duncan:1996xy}
  A.~Duncan, E.~Eichten and H.~Thacker,
  \emph{Electromagnetic Splittings 
        and Light Quark Masses in Lattice QCD}, 
  \emph{Phys.\ Rev.\ Lett.}\ {\bf 76} (1996) 3894
  [{\tt hep-lat/9602005}].
  %%CITATION = HEP-LAT 9602005;%% 
% 
 \bibitem{Yamada} 
  N.~Yamada, T.~Blum, M.~Hayakawa and T.~Izubuchi,   
   \emph{Electromagnetic properties of hadrons with 
         two flavor dynamical domain wall fermions}, 
   in proceedings of 
   \emph{XXIIIrd International Symposium on Lattice Field Theory}, 
  PoS(LAT2005)092. 
\end{thebibliography}
\end{document}